\documentstyle[twoside,fleqn,npb,epsfig]{article}    
\newcommand{\AmS}{{\protect\the\textfont2    
  A\kern-.1667em\lower.5ex\hbox{M}\kern-.125emS}}    
    
\title{Skewed distributions fixed by diagonal partons at small $x, \xi$ and $\gamma^* p   
\rightarrow Vp$ at HERA}    
    
\author{K.~Golec-Biernat\address{H.~Niewodniczanski Institute of Nuclear Physics,   
ul.~Radzikowskiego 152, Krakow, Poland}, {\underline{A.D.~Martin}}\address{Department  
of Physics, University of Durham, DH1 3LE, United Kingdom},  
M.G.~Ryskin\address{St.~Petersburg Nuclear Physics Institute, Gatchina, St.~Petersburg,  
188350, Russia} and A.G.~Shuvaev$^{\rm c}$}    
    
\begin{document}    
    
\begin{abstract}     
We show that the skewed parton distributions are completely determined at small $x$ and   
$\xi$ by the conventional diagonal partons.  We study the application to diffractive vector   
meson production at HERA.  
\end{abstract}    
    
\maketitle    
    
\section{Introduction}  
Data are becoming available for processes which are described by off-diagonal (or skewed)   
parton distributions.  A relevant example is diffractive vector meson production at HERA,   
$\gamma^* p \rightarrow Vp$ with $V = \rho, J/\psi$ or $\Upsilon$, where at high   
$\gamma^* p$ c.m.~energy, $W$, the cross section is dominated by the two-gluon exchange   
diagram  
\begin{equation}  
\left . \frac{d \sigma}{dt} (\gamma^* p \rightarrow Vp) \right |_{t = 0} \; = \; \ldots [x_2 g   
(x_1, x_2; \mu^2)]^2  
\label{eq:1}  
\end{equation}  
where $g$ is the off-diagonal ($x_1 \neq x_2$) gluon distribution with  
\begin{eqnarray}  
x_1 & = & (Q^2 + M_{q\bar{q}}^2)/W^2, \nonumber \\  
x_2 & = & (M_{q\bar{q}}^2 - M_V^2)/W^2 \: \ll \: x_1,  
\label{eq:1a}  
\end{eqnarray}  
see ref.~\cite{MR}.  $M_{q\bar{q}}$ is the mass of the $q\bar{q}$ system produced by a   
photon of virtuality $Q^2$.  The relevant scale is $\mu^2 = z (1 - z) Q^2 + k_T^2 + m_q^2$   
where $z, 1 - z$ and $\pm \mbox{\boldmath $k$}_T$ specify the momenta of the $q$ and   
$\bar{q}$.  The quadratic dependence on $g$ in (\ref{eq:1}) shows that these data may offer   
a sensitive constraint on the gluon.  Indeed our aim is to show that the off-diagonal   
distributions are fixed by the conventional diagonal partons, so that the data can, in principle,   
be included in a global parton analysis.  
  
\section{Ji's \lq symmetrized\rq~distributions}  
  
We shall use the \lq\lq off-forward\rq\rq~distributions $H (x, \xi) \equiv H (x, \xi, t, \mu^2)$   
with support $-1 \leq x \leq 1$ introduced by Ji \cite{JI}, with the minor difference that the   
gluon $H_g = x H_g^{\rm Ji}$ \cite{GM}.  They depend on the momentum fractions  
\begin{equation}  
x_{1,2} \; = \; x \: \pm \: \xi  
\label{eq:2}  
\end{equation}  
carried by the emitted and absorbed partons at each scale $\mu^2$ and on the momentum   
transfer variable $t = (p - p^\prime)^2$.  The variables $t$ and $\xi$ do not change as we   
evolve the distributions up in the scale $\mu^2$.  In the limit $\xi \rightarrow 0$ they reduce   
to the conventional parton distributions  
\begin{eqnarray}  
H_q (x, 0) & = & \left \{ \begin{array}{lcr}  
q (x) & {\rm for} & x > 0 \\  
- \bar{q} (-x) & {\rm for} & x < 0, \end{array} \right .  \\  
H_g (x, 0) & = & xg (x), \nonumber  
\label{eq:3}  
\end{eqnarray}  
and satisfy DGLAP evolution.  In the limit $\xi \rightarrow 1$ they obey ERBL evolution.  If   
we consider $H_q$ at arbitrary values of $\xi$, then for $x > \xi$ and $x < - \xi$ we have   
DGLAP-like evolution for quarks and antiquarks respectively, while for $- \xi < x < \xi$ we   
have ERBL-like evolution for the emitted $q\bar{q}$ pair.  
  
On account of the $x_1 \leftrightarrow x_2$ symmetry the distributions $H_q, H_g$ are   
symmetric in $\xi$.  We also have symmetry relations in terms of the $x$ variable  
\begin{eqnarray}  
H_q^{NS} (x, \xi) & = & H_q^{NS} (-x, \xi), \nonumber \\  
H_q^S (x, \xi) & = & - H_q^S (-x, \xi), \nonumber \\  
H_g (x, \xi) & = & H_g (-x, \xi). \nonumber  
\end{eqnarray}  
where the superscripts $S$ and $NS$ denote singlet and non-singlet quarks respectively.  
  
\section{$H (x, \xi)$ in terms of conformal moments}  
  
The conformal moments\footnote{Conformal moments were introduced in \cite{ER} for $\xi   
= 1$, and in \cite{O} for $\xi \neq 1$; see also \cite{BFKL}.} of the off-diagonal   
distributions,  
\begin{equation}  
\fbox{$O_N (\xi, \mu^2) \; = \; \displaystyle{\int_{-1}^1} dx R_N (x_1, x_2) H (x, \xi),$}  
\label{eq:4}  
\end{equation}  
are not mixed by evolution  
\begin{equation}  
O_N (\xi, \mu^2) \; = \; O_N (\xi, \mu_0^2) \left ( \frac{\mu^2}{\mu_0^2} \right   
)^{\gamma_N},  
\label{eq:5}  
\end{equation}  
where $\gamma_N$ are the same anomalous dimensions as for diagonal partons.  The $R_N$   
are known polynomials of degree $N$  
\begin{equation}  
R_N \; = \; \sum_{k = 0}^N \left ( \begin{array}{c} N \\ k \end{array} \right ) \left (   
\begin{array}{c} N + 2p \\ k + p \end{array} \right ) x_1^k x_2^{N - k}  
\label{eq:6}  
\end{equation}  
with $p = 1,2$ for quarks and gluons respectively. The $O_N$ reduce to the usual moments   
in the limit $\xi \rightarrow 0$.  For example for quarks  
\begin{equation}  
O_N \rightarrow M_N \; = \; \int_0^1 x^N q(x) dx,  
\label{eq:7}  
\end{equation}  
up to a normalizing factor $R_N (1, 1)$.  
  
The crucial step is to find the inverse relation to (\ref{eq:4}).  That is to reconstruct $H (x,   
\xi)$ from a knowledge of the conformal moments.  The result, due to Shuvaev \cite{SHUV},   
is  
\begin{equation}  
\fbox{$H (x, \xi) \; = \; \displaystyle{\int_{-1}^1} dx^\prime K (x, \xi; x^\prime) f (x^\prime)   
$}  
\label{eq:8}  
\end{equation}  
where the kernel $K$ is a known integral \cite{SHUV,SGMR} and $f$ is the Mellin   
transform  
\begin{equation}  
f (x^\prime) \; = \; \int \frac{dN}{2 \pi i} (x^\prime)^{-N} O_N (\xi) / R_N (1, 1).  
\label{eq:9}  
\end{equation}  
$f$ reduces to the diagonal distribution for $\xi^2 \ll 1$.  This follows since \cite{JI}  
\begin{eqnarray}  
O_N (\xi)  & = & \sum_{k = 0}^{[(N + 1)/2]} O_{Nk} \xi^{2k} \nonumber   
\\  
& \simeq & O_{N0} \; = \; O_N (0) = M_N R_N (1, 1)  
\label{eq:10}  
\end{eqnarray}  
for small $\xi^2$.  So the off-diagonal distribution $H$ is completely determined in terms of   
the diagonal distribution $f$ via (\ref{eq:8}).  
  
\section{A good small $x, \xi$ approximation}  
  
We can simplify (\ref{eq:8}) further if we assume that the diagonal partons have the form  
\begin{equation}  
xq (x) \; = \; N_q x^{- \lambda_q}, \quad xg (x) \; = \; N_g x^{- \lambda_g}  
\label{eq:11}  
\end{equation}  
for very small $x$.  Then the $x^\prime$ integration can be performed analytically and  
\begin{equation}  
H_i (x, \xi) \; = \; \xi^{- \lambda_i - p} F_i \left ( \frac{x}{\xi} \right ) \nonumber  
\label{eq:11a}  
\end{equation}  
with $p = 1, 0$ for $i = q, g$ respectively.  A full set of results for the off-diagonal/diagonal   
ratios,   
\begin{equation}  
R_i (x, \xi) \; = \; H_i (x, \xi) / H_i (x + \xi, 0),  
\label{eq:11b}  
\end{equation}  
can be found in \cite{SGMR}.  There, the ratios $R_q^{NS,S}$ and $R_g$ are plotted as   
functions of $x/\xi$ for different values of $\lambda_i$. The scale dependence of the   
off-diagonal distributions, $H_i (x, \xi)$ of (\ref{eq:11a}), and hence of the $R_i$, is hidden   
in the $\mu^2$ dependence of the $\lambda_i$.  Both $\lambda_g$ and $\lambda_q$ increase   
with increasing $\mu^2$.  
  
\section{Application to $\gamma^* p \rightarrow Vp$}  
  
The value of the ratio for the gluon distribution at $x = \xi$ is relevant for diffractive vector   
meson production, $\gamma^* p \rightarrow Vp$, at high energies, see (\ref{eq:1a}).  This   
ratio is given by\footnote{This answer checks with the values of the ratio obtained by direct   
evolution of the off-diagonal and diagonal gluons in \cite{MR}.}  
\begin{equation}  
R_g (x = \xi) \; = \; \frac{2^{2 \lambda_g + 3}}{\sqrt{\pi}} \frac{\Gamma\!\left (\lambda_g +   
\frac{5}{2} \right )}{\Gamma (\lambda_g + 4)}.  
\label{eq:12}  
\end{equation}  
The cross section formula (\ref{eq:1}) may then be expressed in terms of the conventional   
diagonal gluon distribution $g$,  
\begin{equation}  
\left . \frac{d \sigma}{dt} (\gamma^* p \rightarrow Vp) \right |_{t = 0} \; = \; \ldots \left [R_g   
x_1 g (x_1, \mu^2) \right ]^2,  
\label{eq:13}  
\end{equation}  
where all the off-diagonal effects are contained in the known (enhancement) factor $R_g^2$.    
Of course to calculate the cross section properly we must use the unintegrated gluon   
distribution and integrate over the transverse momenta of the exchanged gluons and of the   
$q$ and $\bar{q}$ forming the vector meson.  
  
To obtain the scale dependence of $R_g$, we first obtain the $\mu^2$ dependence of   
$\lambda_g$ of (\ref{eq:11}) from the behaviour of the gluon found in the global parton   
analyses.  For example, the MRST partons \cite{MRST} have $\lambda_g = 0.205$ and 0.38   
at $\mu^2 = 4$ and 100~GeV$^2$ respectively.  The appropriate scale for the diffractive   
process $\gamma^* (Q^2) p \rightarrow V (q\bar{q}) p$ is $\mu^2 \simeq m_q^2 + Q^2/4$.    
In this way, for diffractive $J/\psi$ and $\Upsilon$ photoproduction at HERA we find that the   
off-diagonal enhancement, $R_g^2$, is $(1.15)^2$ and $(1.32)^2$ respectively.  However,   
for $\Upsilon$ photoproduction, $x$ is not sufficiently small $(\sim 0.01)$ and we have to   
improve the assumption made in (\ref{eq:11}).  If we take $xg \sim x^{- \lambda_g} (1 -   
x)^6$ and perform the $x^\prime$ integration in (\ref{eq:8}) numerically, then we find an   
enhancement of $(1.41)^2$ for $\Upsilon$ photoproduction \cite{TEUB}.  Moreover for   
$\rho$ electroproduction it is found \cite{MRT} that the enhancement due to off-diagonal   
effects of the $\gamma^* p \rightarrow \rho p$ cross section $d \sigma/d Q^2$, at the largest   
$Q^2$ of the HERA data, is more than a factor 2, which is just the enhancement needed to 
ensure a perturbative QCD description of the data.  
  
\section{Discussion}  
  
The main conclusion is embodied in eqs. \linebreak (\ref{eq:8})--(\ref{eq:10}).  That is the   
skewed distribution $H (x, \xi)$, at any scale, is fully determined at small $x, \xi$ by   
knowledge of the diagonal parton distribution, at the same scale.  
  
To be sure of this result we have checked that the analytic continuation of the conformal   
moments $O_N$ in $N$ is allowed \cite{SGMR}.  A second consideration is that, from a   
formal point of view, we may add to the off-diagonal distribution any function which exists   
only in the ERBL-like region, $|x| < \xi$.  In \cite{SGMR} we show such a contribution is   
negligible $O (\xi^2)$ at small $\xi$.  So far our distributions allow the calculation of the   
imaginary part of the amplitude for the process.  At small $x$ and $\xi$ it turns out that the   
real part may be calculated easily using a dispersion relation in the c.m.~energy squared,   
$W^2$, and that the amplitude  
\begin{equation}  
A \; = \; i {\rm Im} A \: \frac{1 + e^{- i \pi \lambda}}{1 + \cos \pi \lambda},  
\label{eq:14}  
\end{equation}  
where $A \propto (W^2)^\lambda$.  Finally we note that our result remains valid at NLO,   
since there is no conformal mixing for $\xi^2 \ll 1$.  
  
We conclude that, at small $x, \xi$, the skewed distributions $H (x, \xi; \mu^2)$ are   
completely known in terms of conventional partons.  Thus data for processes which are   
described by such distributions can, in principle, be included in a conventional global analysis   
to better constrain the low $x$ behaviour of the partons. \\  
  
\noindent {\bf Acknowledgements}  
  
\medskip  
We thank Max Klein and Johannes Bl\"{u}mlein for their efficient organization of DIS99,   
and the Royal Society and the EU Fourth Framework Programme `Training and Mobility of  
Researchers', Network `QCD and the Deep Structure of Elementary Particles', contract  
FMRX-CT98-0194 (DG 12-MIHT) for support.

\end{document}